# Annals of Library and Information Studies: A bibliometric analysis of the journal and a comparison with the top library and information studies journals in Asia and worldwide (2011–2017)

Juan Jose Prieto-Gutierrez & Francisco Segado-Boj

**PREPRINT**
**Paper accepted for publication in The Serials Librarian**

**Abstract**
This paper presents a thorough bibliometric analysis of research published in *Annals of Library and Information Studies* (ALIS), an India-based journal, for the period 2011–2017. Specifically, it compares this journal's trends with those of other library and information science (LIS) journals from the same geographical area (India, and Asia as a whole) and with the 10 highest-rated LIS journals worldwide. The source of the data used was the multidisciplinary database Scopus.

To perform this comparison, ALIS' production was analyzed in order to identify authorship patterns; for example, authors' countries of residence, co-authorship trends, and collaboration networks. Research topics were identified through keyword analysis, while performance was measured by examining the number of citations articles received.

This study provides substantial information. The research lines detected through examining the keywords in ALIS articles were determined to be similar to those for the top LIS journals in both Asia and worldwide. Specifically, ALIS authors are focusing on metrics, bibliometrics, and social networking, which follows global trends.

Notably, however, collaboration among Asia-based journals was found to be lower than that in the top-indexed journals in the LIS field. The results obtained present a roadmap for expanding the research in this field.

**Introduction and literature review**
The field of library and information science (LIS) has experienced significant growth in recent years, with the number of titles covered by Scopus increasing by 30% over the last decade.

Analysis of scholarly production through bibliometrics facilitates interpretation of the structures of and trends in particular disciplines[1]. Such bibliometric analysis affords identification of the characteristics of a research field, and can contribute to revealing future research topics[2]. This analysis can be implemented at different levels to suit the nature of the actors under investigation; for instance, at the national (countries), individual (authors), or institutional (universities and research centers) level[3]. Further, bibliometrics can serve to identify collaboration patterns among authors and, through keyword analysis, thematic overviews of a scientific discipline[4].

Along with allowing researchers to investigate authorship and collaboration, bibliometric analyses can be implemented at a conceptual or intellectual level. As a consequence, studies have found that geographical distance, specialization patterns, and cultural proximity are significant factors that positively affect scientific collaboration between regions[5]. Further, it has also shown that, in regard to international collaboration, advanced knowledge and technologies are of primary interest[6].

The current study seeks to analyze the LIS discipline in terms of three geographical levels: local, regional, and global. Dividing the research field into such separate levels has previously been conducted in many papers that focused on the "state of LIS." For example, from a local point of view, there are studies that have focused only on Spain[7] or India[8]; in terms of regional research, there are papers that have focused only on Asia[9] or Eastern European countries[10]; and in regard to global levels, there has been research such as a content analysis of articles from the 10 highest-rated LIS journals[11], and a study that reviewed 217 LIS journals for a content analysis of LIS literature[12].

For this study, we have chosen a journal of Indian origin, *Annals of Library and Information Studies* (ALIS), as a representative LIS journal, and sought to compare its output and author statistics with similar journals from Asia and worldwide. ALIS is a leading quarterly LIS journal, which has risen since the fourth quartile in Scimago Journal Rank in 2012 to the second quartile in 2017 (Scimago Journal Rank). It has been published since 1954 by the CSIR-National Institute of Science Communication and Information Resources (CSIR-NISCAIR), formerly the Indian National Scientific Documentation Center. During the period analyzed in this study (2011–2017). ALIS achieved the highest position in the Scimago Journal Rank (SJR) for the LIS category, not only for India (2013, 2014, 2015), but also for the entire Asiatic area (2014 and 2015). This means that ALIS is a significant journal in terms of providing relevant research in its field and geographical area.

Specifically, the present study presents a bibliometric analysis of the articles published by ALIS from 2011 to 2017. The results are compared to those of a previous analysis of ALIS articles for the period 2002–2010[13]. Such a comparison can serve to identify trends in the evolution of the field by highlighting the most relevant bibliometrics indexes. The results are also compared with the top LIS journals in Asia and worldwide.

This approach allows us to identify specific patterns in LIS production in the Indian and Asian regions, and to compare the trends with those of the leading journals in the field.

**Objectives of the study**
The main objective of the present study is to obtain objective information regarding papers published by ALIS, and its contributing authors, over the last seven years (2011–2017), and to compare this information with that of other Asia-based journals and the 10 highest-rated LIS journals in the Scopus database. Thus, it seeks to identify specific trends at the journal, regional, and international levels.

The specific research objective is:

- To research bibliometric data for ALIS in order to identify the year-wise distribution of the journal's articles, the journal's authorship pattern, and the journal's most productive authors.

To characterize ALIS' position in the international context, which would serve to identify benchmarks for determining the journal's overall status in the field, we must establish:
- The 20 most frequently used author keywords during 2011–2017 for ALIS, among all LIS journals in Asia, and among the 10 highest-rated LIS journals worldwide.
- The countries of residence of the corresponding authors for the three levels.
- Total citations per country for the three levels.

**Methodology**
For this research, data were sourced from the SJR and the Scopus database, using the category "library and information science," and focusing on the period 2011–2017. The Scopus database was chosen because studies have shown it to have better geographic and thematic coverage when compared with other databases[14]. Regarding geographic distribution, we worked at the macro level (country, regional, and worldwide ranks), and we used a set of indicators (published articles, authorship collaboration, keyword analysis) to derive complementary data. This methodologic approach permitted analysis of librarians, publishers, and researchers.

We used a three-step approach for data collection and comprehensive evaluation of the LIS field. First, we analyzed ALIS; then, all LIS journals in the Asia region; and third, the 10 highest-rated LIS journals worldwide.

The results have the potential to provide, across the three levels, insights regarding current research interests, avenues of future research, the main countries that contribute to the field, and geographical affiliations.

Such research is usually performed through a cycle of searching keywords, searching the literature that is returned[15], designing mind maps to structure the results[16], and completing the analysis in a manner that allows the research results to be published. A similar approach was followed in the present research.

The data were retrieved on January 1, 2019, from the Scopus database. All the information was collected, organized, and analyzed using the Bibliometrix package for R software[17].

**Results**

Table 1 shows the production statistics for ALIS articles for each year of the study period.

Table 1. Year-wise distribution of articles in *Annals of Library and Information Studies* for 2011–2017.

| Year (Vol.) | Issue no. | | | | Total no. of articles | Cumulative number of articles | % of articles |
| --- | --- | --- | --- | --- | --- | --- | --- |
| | No. 1 | No. 2 | No. 3 | No. 4 | | | |
| 2011 - Vol. 58 | 10 | 10 | 9 | 7 | 36 | 36 | 15.06% |
| 2012 - Vol. 59 | 6 | 6 | 9 | 8 | 29 | 65 | 12.13% |
| 2013 - Vol. 60 | 9 | 9 | 9 | 0 | 27 | 92 | 11.29% |
| 2014 - Vol. 61 | 9 | 8 | 11 | 10 | 45 | 137 | 18.82% |
| 2015 - Vol. 62 | 6 | 7 | 9 | 16 | 38 | 175 | 15.89% |
| 2016 - Vol. 63 | 10 | 8 | 8 | 6 | 32 | 207 | 13.38% |
| 2017 - Vol. 64 | 10 | 6 | 6 | 10 | 32 | 239 | 13.38% |

Table 1 also shows summary statistics for the data, and the annual scientific production of articles by year, volume, and issue. The year (and volume) with the highest number of total articles was 2014 (vol. 61) with 45, while the lowest was 2013 (vol. 60), with 27. Over the entire period, the annual percentage growth rate was −1.94%.

Comparing Table 1 with a table created by Jena, Swain, and Sahoo (2012) shows a huge rise in the number of articles published during the second period (2011–2017).
Over the 15 years in question, the journal published 454 documents. Figure 1 provides a graphical description of the upwards trend in the quantity of articles published.

In the early years of 2002 and 2003, the journal published approximately 20 articles per year, and by 2017 was publishing approximately 35 papers each year. It is worth mentioning that 2013, with just 27 documents published, represents a temporary break in the ascending trend.

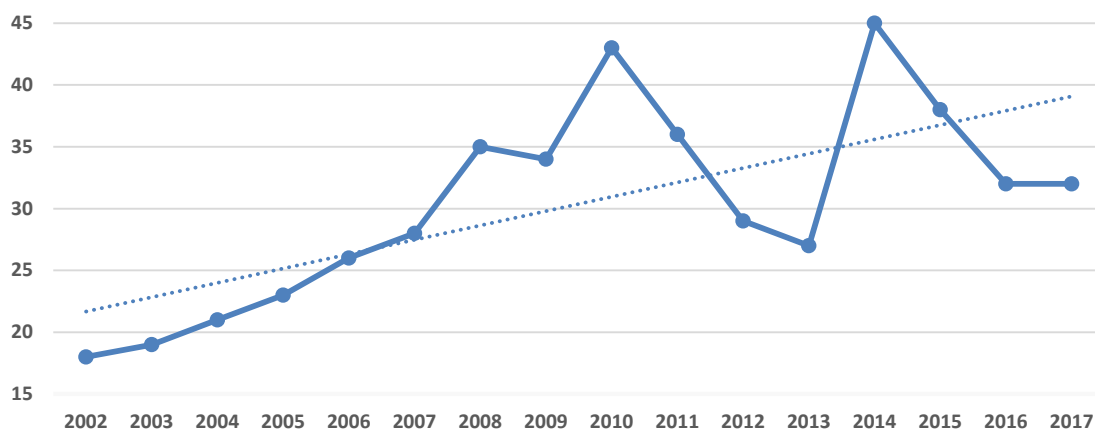

Figure 1. Publishing trend for *Annals of Library and Information Studies* from 2002 to 2017.

Table 2. Authorship pattern.

| SI No. | Rank | Authorship pattern | No. of contributions (2011–2017) | % of contributions | No. of contributions (2002–2010) | % of contributions |
|---|---|---|---|---|---|---|
| 1 | 2 | Single | 86 | 36.28% | 80 | 32.38% |
| 2 | 1 | Two | 103 | 43.45% | 117 | 47.36% |
| 3 | 3 | Three | 38 | 16.03% | 43 | 17.4% |
| 4 | 4 | > Three | 10 | 4.21% | 7 | 2.83% |

Table 2 indicates that most papers (103; 43.54%) were written by two authors, followed by single authors (86; 36.28%), three authors (38; 16.03%), and more than three authors (10; 4.21%).

Next, Table 3 outlines the 10 highest-contributing authors and the quantity of papers they authored or co-authored.

Table 3: Top 5 contributors and their productivity for 2011–2017.

| SI No. | Authorship pattern | No. of contributions |
|---|---|---|
| 1 | Sen, B. K | 17 |
| 2 | Garg, K. C. | 8 |
| 3 | Gupta, B. M. | 8 |
| 4 | Dutta, B. and Pujar, S. M. | 6 |
| 5 | Four authors | 5 each |

The bibliometric analysis of the 239 papers showed that, from this period, 327 different authors published papers (0.731 documents per author). But the authors appearances were 454 (authors per document 1.37)

As can be seen in this data, B. K. Sen dominates the list, as he is the most prolific author, publishing 17 papers between 2011 and 2017. Meanwhile, K. C. Garg and B. M. Gupta are *ex aequo* over the period, with both producing eight articles each.

Forty-five individuals published single-authored documents (with some publishing more than one single-authored paper), and 282 contributed to multi-authored documents. These data indicate that over 85% of authors performed their research collaboratively. It should be noted that the co-authors per document index was 1.9, meaning that the collaboration index was 1.84.

**Collaboration patterns for the three most-productive authors**

In the next phase, we analyzed the collaboration patterns for the three most-productive authors in ALIS using VOSviewer[18]; a collaboration map is presented in Figure 2.

The map features several different components, including circles of varying sizes, node-networked relationship clustering (color and proximity), and text (featuring the authors' names). For each author, the font size of the item's label and the size of the item's circle are based on the weight of the item. Gupta, as the most active author in recent years, is given the largest circle and label font size.

For each of the authors, the number of bibliographic coupling links was calculated. In Figure 2, authors who are attributed similar circle colors are considered a cluster (meaning they have a close collaboration).

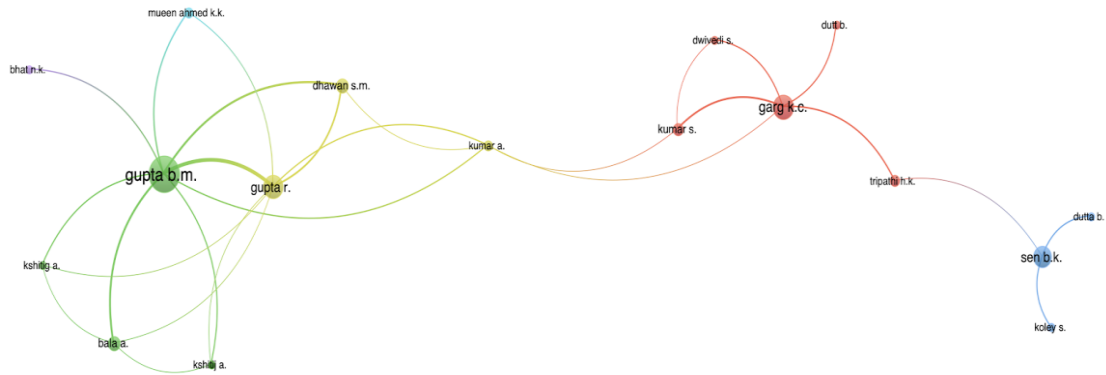

Figure 2. Network visualization of the author collaboration of ALIS' three most-productive authors for 2011–2017.

The collaboration network of ALIS' three most-productive researchers shows that Gupta is the researcher with the most coauthors, and the best connected. Garg and Sen have lower numbers of coauthors

When the network analysis is extended to the journal level, the results are quite similar. Sen's collaboration network remains the same, which indicates that almost all of his publications in ALIS are with the same collaborators. In contrast, Garg shows fewer connections than in the general network. Further, Gupta does not even appear in this collaboration network.

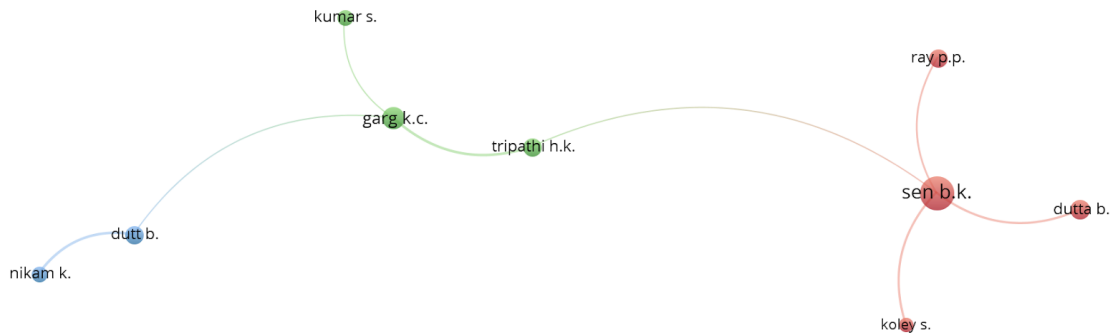

Figure 3. Network visualization of the author to the ALIS' journal level

**Outstanding research topics**

Our next step was to select for examination journals from other major geographic zones in order to compare the results with those for ALIS. For this, we chose to consider the scientific output of LIS journals based in Asia and the 10 highest-rated LIS journals worldwide.

Consequently, we examined a set of 20 core LIS journals for the 2011–2017 period. Specifically, we followed the SJR for 2017 to select the 10 highest-rated LIS journals based in Asia and worldwide, respectively.

Our analysis focused on the following: number of publications (Table 4), frequently used author keywords (Table 5), corresponding authors' countries of residence (Table 6), and the geographic distribution of contributions and citations per country (Table 7).

Table 4 displays the two lists of journals (Asia and worldwide) and their respective total number of publications for the last seven years. The 10 highest-rated journals worldwide produced a cumulative total of 6,018 articles, while the 10 highest-rated Asia-based journals produced 5,139. In Asia, the *Journal of Information and Computational Science* ranked first with 3,090 papers, approximately five times more than the second-placed *Journal of Information Science and Engineering*, and accounting for 60% of all articles in the top 10.

Meanwhile, at the worldwide level the *Journal of Chemical Information and Modeling* published the most articles (1,992), and *Communications in Information Literacy* published the lowest (131) over the last seven years.

Table 4. The 10 highest-rated journals in Asia and worldwide for 2011–2017.

|   | The 10 highest-rated journals in the Asia region | TP | The 10 highest-rated journals worldwide | TP |
|---|---|---|---|---|
| 1 | DESIDOC Journal of Library and Information Technology | 363 | Information Systems Research | 392 |
| 2 | Malaysian Journal of Library and Information Science | 150 | Scientific Data | 812 |
| 3 | Annals of Library and Information Studies | 239 | Information Communication and Society | 620 |
| 4 | Journal of Information Science and Engineering | 632 | Journal of Informetrics | 634 |
| 5 | Journal of Educational Media and Library Science | 75 | Journal of Information Technology | 207 |
| 6 | Pakistan Journal of Information Management and Libraries | 45 | Communications in Information Literacy | 131 |
| 7 | Library and Information Science | 124 | European Journal of Information Systems | 287 |
| 8 | Journal of Digital Information Management | 383 | College and Research Libraries | 333 |
| 9 | Journal of Information and Computational Science | 3,090 | International Journal of Information Management | 610 |
| 10 | Progress in Informatics | 38 | Journal of Chemical Information and Modeling | 1,992 |
|   | TOTAL | 5,139 |  | 6,018 |

*TP*: Total number of publications.

Keywords provide important information regarding research trends and limits, revealing research fields and focuses of interest[19]; however, their use is relatively recent[20].

The 5,139 papers produced by the 10 highest-rated Asian journals contained 15,061 unique author keywords.
The 20 most-frequently used keywords are displayed in Table 5. Every keyword identified was counted and ranked within each of the seven one-year intervals in the 2011–2017 period. We compared the keywords presented in ALIS articles with those in articles published in the 10 highest-rated LIS journals published in Asia and worldwide, respectively.

Table 5. Twenty most-frequently used author keywords during 2011–2017.

| ALIS keywords | TP | Asia-based LIS journal keywords | TP | Ten highest-rated LIS journals worldwide keywords | TP |
| --- | --- | --- | --- | --- | --- |
| Scientometrics | 20 | Genetic algorithm | 72 | Social media | 200 |
| India | 19 | Wireless sensor networks | 66 | Bibliometrics | 72 |
| Bibliometrics | 17 | Bibliometrics | 56 | Citation analysis | 70 |
| Nigeria | 9 | Data mining | 56 | ICTS | 48 |
| E-resources | 8 | Support vector machine | 54 | Twitter | 46 |
| Academic libraries | 7 | Scientometrics | 48 | H-index | 45 |
| Information literacy | 7 | Cloud computing | 47 | Digital divide | 44 |
| Citation analysis | 6 | Particle swarm optimization | 46 | Knowledge management | 44 |
| Impact factor | 6 | Face recognition | 43 | Research evaluation | 43 |
| LIS journals | 6 | India | 43 | Politics | 40 |
| Open access | 6 | Information literacy | 42 | Privacy | 40 |
| Web 2.0 | 6 | Clustering | 39 | Facebook | 39 |
| Colon classification | 5 | Cognitive radio | 35 | Web 2.0 | 39 |
| Consortia | 5 | Wireless sensor network | 35 | Cloud computing | 38 |
| Electronic resources | 5 | Citation analysis | 34 | Internet | 38 |
| Internet | 5 | Feature extraction | 34 | Social movements | 36 |
| Sri Lanka | 5 | Optimization | 34 | Trust | 36 |
| University libraries | 5 | Classification | 33 | Innovation | 34 |
| Bangladesh | 4 | Academic libraries | 31 | Social networks | 32 |
| Citations | 4 | Image segmentation | 30 | Case study | 31 |

*TP:* Total number of publications.

The 20 keywords for each level reflect the respective content and research methods of the articles in question, and reveal the LIS-related research priorities of international and community regions. Further, through considering regional and cultural patterns, the keywords also reveal limitations in study areas (such as through the absence of certain keywords).

For ALIS, "scientometrics," "bibliometrics," and "e-resources" attracted most attention. Meanwhile, the Asia region, "genetic algorithm," "wireless sensor networks," and "bibliometrics" were the main focus, and for the 10 highest-rated journals worldwide, "social media," "bibliometrics," and "citation analysis" were emphasized.
Furthermore, content analysis of the keywords indicated the varying research directions of the LIS field. For example, in the 10 highest-rated journals worldwide, social media, social movements, and social networking (Twitter and Facebook) were the main issues.
Meanwhile, the Asia region is dominated by Chinese researchers who are largely interested in technology related to health aspects. At this level, among LIS articles, the topics of metrics and data analysis are the main focus of the LIS scientific field.
"Bibliometrics" was the most common keyword in the three studied areas, situated in second or third place for each level.

We also found some keywords that were representative of the geographical study area, such as "India," "Nigeria," and "Sri Lanka" in ALIS, and "India" in the Asia-based journals; the appearance of "India" in both demonstrates the close relationship among Asian countries in regard to LIS research.

Table 6. Corresponding authors' countries of residence.

|   | ALIS | TP | SCP | MCP | Asia region | TP | SCP | MCP | Ten highest-rated journals worldwide | TP | SCP | MCP |
|---|------|----|-----|-----|-------------|----|-----|-----|--------------------------------------|----|-----|-----|
| *1* | *India* | *154* | *151* | *3* | China | 2,473 | 2,342 | 131 | USA | 1,250 | 997 | 253 |
| 2 | Nigeria | 17 | 16 | 1 | *India* | 451 | 444 | 7 | UK | 398 | 278 | 120 |
| 3 | Sri Lanka | 8 | 8 | 0 | Taiwan | 314 | 297 | 17 | Germany | 334 | 240 | 94 |
| 4 | Bangladesh | 6 | 5 | 1 | Korea | 97 | 86 | 11 | China | 252 | 150 | 102 |
| 5 | Tanzania | 2 | 2 | 0 | Japan | 47 | 45 | 2 | Italy | 181 | 132 | 49 |
| 6 | Fiji | 1 | 1 | 0 | Pakistan | 43 | 37 | 6 | Canada | 125 | 96 | 29 |
| 7 | Iran | 1 | 1 | 0 | Malaysia | 42 | 35 | 7 | Korea | 114 | 69 | 45 |
| 8 | South Africa | 1 | 1 | 0 | USA | 41 | 27 | 14 | Spain | 114 | 82 | 32 |
| 9 | Uganda | 1 | 1 | 0 | Iran | 29 | 28 | 1 | Australia | 110 | 69 | 41 |
| 10 | USA | 1 | 1 | 0 | Nigeria | 29 | 26 | 3 | France | 99 | 61 | 38 |
|   |   |   |   |   |   |   |   |   | *17. India* | *55* | *40* | *15* |

*TP:* Total number of publications
*SCP*: Single-country publications
*MCP*: Multiple-country publications

Next, information regarding geographic distribution was retrieved by examining corresponding authors' affiliations. Table 6 shows, for 2017, the 10 most-productive countries for each level of our analysis. The three lists of countries are ranked based on their scientific output.

As in the previous table, we have three columns; the first represents the countries of residence of the corresponding authors of ALIS articles, the second presents the equivalent data for the Asia-region journals, and the third presents that same data for the 10 highest-rated journals worldwide.

Considering the third column, the USA dominates (1,250 articles), with three times as many articles as the second-placed UK (398). China is located in fourth position (252), and India is situated in 17[th] position (55 articles), which can be considered an acceptable position in the field.

For Asia (the second column), China (2,473) is a powerful research force in the field, producing approximately five times as many articles as second-placed India (451). The USA (41) does not have strong representation in this geographical area, and is consequently located in eighth position, with a comparatively low output.

Although China contributes markedly to Asian research, it performs approximately average on other indexes; for example, for ALIS, India is the clear leader, accounting for 80% of all publications.

The same table affords analysis of international collaboration. This can be performed by counting the number of articles written by authors from the same country (SCP; such publications represent intra-country collaboration), and the number of articles featuring authors from different countries (MCP; such publications represent international collaboration).

All countries across all three columns published more single-country publications than internationally collaborative publications.

In ALIS, approximately 98% of the articles (151) were published by domestic authors; this indicates that the journal is primarily of local interest.

For the Asia region, China ranked first in all indicators, such as total SCP (94%). Among the Indian journals, over 98% of the articles were published by domestic authors, as evidenced by the contributors to ALIS.

Analysis of the 10 highest-rated journals worldwide shows that the USA accounted for 20% of the internationally collaborative publications. Korea and China presented the most international collaborative publications with 40% of all its papers; Australia and France ranked second with 38%.

It is clear that single-country publications dominate LIS research, which is in contrast to other scientific areas, in which international collaboration is a general trend[21].

Table 7. Total citations per country for 2011–2017.

|   | ALIS | TC/C | AAC | Asia region | TC/C | AAC | Ten highest-rated journals worldwide | TC/C | AAC |
|---|---|---|---|---|---|---|---|---|---|
| 1 | **India** | **365** | **2.37** | China | 4,392 | 1.77 | USA | 30,750 | 24.60 |
| 2 | Bangladesh | 19 | 3.17 | **India** | **976** | **2.16** | UK | 8,445 | 21.22 |
| 3 | Nigeria | 18 | 1.06 | Taiwan | 770 | 2.45 | China | 6,276 | 24.90 |
| 4 | Sri Lanka | 14 | 1.75 | Korea | 345 | 3.55 | Germany | 6,055 | 18.13 |
| 5 | Uganda | 10 | 10 | Malaysia | 251 | 5.97 | Italy | 2,819 | 15.57 |
| 6 | Iran | 3 | 3 | Iran | 125 | 4.31 | Canada | 2,667 | 21.34 |
| 7 | Tanzania | 3 | 1.5 | Pakistan | 80 | 1.86 | Taiwan | 2,558 | 30.09 |
| 8 | Fiji | 1 | 1 | Germany | 50 | 7.14 | Korea | 2,458 | 21.56 |
| 9 | USA | 1 | 1 | Thailand | 47 | 3.91 | Switzerland | 2,370 | 28.90 |
| 10 | South Africa | 0 | 0 | USA | 40 | 0.97 | Spain | 2,289 | 20.08 |
|   |   |   |   |   |   |   | *13. India* | *1,480* | *26.91* |

*TC/C*: Total citations per country
*AAC*: Average number of article citations.

The total citations per country (TC/C) and average number of article citations[22] are shown in Table 7. Both represent means of measuring geographical areas of research. Table 7 shows the number of citations per country for the three levels across the study period; significant differences were found in this regard.

The number of citations is similar to those shown in Table 6 and, in most cases, the order of the countries is the same across the two tables. This suggests a correlation between the corresponding authors' countries and TC/C; this can be explained by the fact that the collaboration networks are mainly composed of authors from the same country.

For the 10 highest-rated LIS journals worldwide, in terms of citations, the USA outnumbers all other countries (30,750), with over three times the number of the next most-cited country, UK, which had 8,445 citations. Three Asian countries are situated in the top 10: China (6,276), Taiwan (2,558), and Korea (2,458). In this ranking, India appears in 13th position (with 1,480), but is ranked with a high average article citation of 26.91, which is the third highest in this regard, above the USA (24.60) and the UK (21.22).

Asia is led by China (4,392), and then India (976); meanwhile, for ALIS articles, India accounted 80% of the citations.

In summary, in the three levels considered, analysis of TC/C revealed a slightly different order to that for corresponding authors' countries of residence, but with India, China, and the USA still far ahead of the other countries.

At the regional and worldwide levels, Chinese research is the most diversified and cited among all countries. India has a similar status at the local and regional levels (behind China)

**Discussion**

As can be noted from Scimago Journal Rank World scientific output in LIS is currently low, but is increasing annually. The regional distribution in this field is concentrated mostly in Europe (50%) and North America (40%); the Asian region only accounts for 5%, but is witnessing growth.

Within the Asia region, India is the top producer, with three notable journals (*DESIDOC, Annals of Library Studies* and *Journal of Digital Information Management*), and has high levels of specialization in the LIS area. However, on a global level it does not have a notable presence.

In the 10 highest-rated LIS journals worldwide, the USA dominates regarding authorship, and China is also a powerful research force in the Asia region, with India ranked second in this latter level.

The statistics for ALIS are very similar to those for the most important LIS magazines in the Asian region, and even the most important in the world.

Among ALIS' greatest strengths is the sustained growth of its output (as shown in Figure 1), and the topics and areas of research interest in its articles. Our analysis of

keywords indicated the research directions of the LIS field, at all three levels. Through this, we concluded that the current interests are technology and, specifically, metrics, bibliometrics, and social networking, which follows global trends (which we determined through analyzing the 10 highest-rated LIS journals worldwide).

The largest number of citations of ALIS articles come from India, which is to be expected. However, expanding the analysis to citations of LIS journals in the Asian region shows India in second place, behind China, but with a higher number of average citations per article than China.

Analyzing the most important journals in the world, we find India in 13$^{th}$ place, with a high average number of citations per article; even higher than the United States, which has the highest number of total citations in the LIS field.

Among ALIS' weaknesses is its very low level of international collaboration, despite the internationalization of science. In the Asia region, approximately 90% of the papers are single-country publications, and this is a general tendency, not a phenomenon restricted to only a few countries. This behavior should change, especially considering that multi-country publications receive more citations than do single countries.

An increase in multi-national collaboration was evident when the 10 highest-rated journals worldwide were considered; here, 30% of the contributions were multi-country publications.

The results of the present research can represent a useful means of performing future diagnostics of ALIS' research capacity. Such information is relevant for situating the LIS field in the regional level and in the world. Additionally, the methodology used in the present research could be applied to analysis of other journals and scientific disciplines, as it can contribute to identifying trends and likely future developments in this regard.

NOTES

[1] Julio Montero-Díaz, Manuel Jesús Cobo, María Gutiérrez-Salcedo, Francisco Segado-Boj, Enrique Herrera-Viedma, "A Science Mapping Analysis of'Communication'WoS Subject Category (1980-2013)." *Comunicar: Media Education Research Journal* 26.55 (2018): 81-91.

[2] Santa Soriano, Alba, Carolina Lorenzo Álvarez, and Rosa María Torres Valdés. "Bibliometric analysis to identify an emerging research area: Public Relations Intelligence—a challenge to strengthen technological observatories in the network society." *Scientometrics* 115.3 (2018): 1591-1614.

[3] Yuh- Shan Ho and James Hartley. "Classic articles in psychology in the Science Citation Index Expanded: a bibliometric analysis." *British Journal of Psychology* 107.4 (2016): 768-780.

[4] Chiu, Wen-Ta, and Yuh-Shan Ho. "Bibliometric analysis of tsunami research." *Scientometrics* 73.1 (2007): 3-17.

[5] Manuel Acosta, Daniel Coronado, Esther Ferrándiz, María León; "Factors affecting inter-regional academic scientific collaboration within Europe: The role of economic distance." *Scientometrics* 87.1 (2011): 63-74.

[21] Olle Persson, Wolfgang Glänzel, and Rickard Danell. "Inflationary bibliometric values: The role of scientific collaboration and the need for relative indicators in evaluative studies." *Scientometrics* 60.3 (2004): 421-432.

[22] Olisah Chijioke, Omobola O. Okoh, and Anthony I. Okoh. "A bibliometric analysis of investigations of polybrominated diphenyl ethers (PBDEs) in biological and environmental matrices from 1992–2018." *Heliyon* 4.11 (2018): e00964.